# Universal growth scheme for entanglement-ready quantum dots

Joanna Skiba-Szymanska[1*], R. Mark Stevenson[1], Christiana Varnava[1,2], Martin Felle[1,2], Jan Huwer[1], Tina Müller[1], Anthony J. Bennett[1], James P. Lee[1,2], Ian Farrer[3,4], Andrey Krysa[4], Peter Spencer[3], Lucy E. Goff[3], David A. Ritchie[3], Jon Heffernan[4], Andrew J. Shields[1]

[1]*Toshiba Research Europe Limited, 208 Science Park, Milton Road, Cambridge CB4 0GZ, UK*
[2]*Cambridge University Engineering Department, 9 J J Thomson Avenue, Cambridge CB3 0FA, UK*
[3]*Cavendish Laboratory, University of Cambridge, J J Thomson Avenue, Cambridge CB3 0HE, UK*
[4] *Department of Electronic and Electrical Engineering, University of Sheffield, Sheffield S1 3JD, UK*

**Abstract**

Efficient sources of individual pairs of entangled photons are required for quantum networks to operate using fibre optic infrastructure. Entangled light can be generated by quantum dots (QDs) with naturally small fine-structure-splitting (FSS) between exciton eigenstates. Moreover, QDs can be engineered to emit at standard telecom wavelengths. To achieve sufficient signal intensity for applications, QDs have been incorporated into 1D optical microcavities. However, combining these properties in a single device has so far proved elusive. Here, we introduce a growth strategy to realise QDs with small FSS in the conventional telecom band, and within an optical cavity. Our approach employs 'droplet-epitaxy' of InAs quantum dots on (001) substrates. We show the scheme improves the symmetry of the dots by 72%. Furthermore, our technique is universal, and produces low FSS QDs by molecular beam epitaxy on GaAs emitting at ~900nm, and metal-organic vapour phase epitaxy on InP emitting at ~1550 nm, with mean FSS 4x smaller than for Stranski-Krastanow QDs.

Entangled photon pair sources based on the radiative decay of the biexciton state within semiconductor quantum dots[1,2] have many potential advantages, including electrical operation[3], sub-Poissonian statistics, potential for integration with conventional optoelectronics, and wide wavelength coverage[4]. Quantum dots (QDs) typically present an energetic fine-structure-splitting (FSS) between the exciton eigenstates much larger than the natural linewidth of the transition[5,6], providing 'which path' information complicating observation of entanglement. The FSS originates from asymmetry in the QD wavefunction caused by variations in crystal strain and composition, and elongation of the QD shape which itself is a natural consequence of the Stranski-Krastanow (S-K) growth technique.

Production of quantum dots with naturally small FSS may be achieved in special cases, by self-assembled growth of InAs quantum dots around 885 nm, where inversion of the wavefunction symmetry is observed[7], or growth on (111) surfaces[8–10], aided by the underlying $C_{3v}$ crystal symmetry. Though these methods are sufficient to enable observation of entangled light from quantum dots[2,8,9], they are unsuitable for entangled light emission in the conventional fibre transmission window around 1550 nm, for which an alternative approach is required.

Here, we demonstrate droplet epitaxy is a viable route to achieving QDs with small FSS, extending to conventional telecom wavelengths. In addition we show the method is suitable for growth on standard (001) surfaces preferred for high quality optical cavity growth. We investigate two material systems: InAs/GaAs and InAs/InP grown by two different growth techniques: MBE and MOVPE producing dots at 900nm and 1550nm. The growth approach relies on formation of metal droplets on the semiconductor substrate followed by crystallisation in an arsenic environment. The density and size of the droplets can be independently controlled with deposition temperature and metal volume respectively. Moreover, since formation of D-E QDs is not strain driven, the shape of the QD is less elongated, and their FSS remains small.

For fibre optic applications, such as quantum communications it is important that the dots emit in the telecom bands. In this respect, InAs/InP QDs are of particular interest, as the reduction in strain compared to InAs/GaAs QDs lengthens the emission wavelength to the conventional telecom band around 1550 nm. InAs QDs on InP can be grown epitaxially by MOVPE or MBE, however achieving low density is far more challenging than on GaAs substrates, without etching of nanomesas[11,12]. In order to reduce the dot densities and shift emission wavelength GaAs interlayers[13] and double capping methods[14] have been proposed. Another problem is formation of single quantum dashes rather than dots directly on InP[12,15]. Employing droplet epitaxy growth technique on InP enabled us to meet the low density criteria at required emission wavelengths necessary for single dot spectroscopy, and also to address the formation of dashes.

In the S-K (Stranski-Krastanow) growth mode of InAs quantum dots, arsenic and indium fluxes are supplied to the substrate at the same time. Initially, growth is two-dimensional, and a highly strained InAs wetting layer is formed. The strain, caused by lattice mismatch between the InAs and the GaAs/InP substrate, increases with the thickness of the wetting layer. Once the wetting layer reaches a critical thickness, typically 1-2 monolayers, it becomes energetically favourable to form QDs, where the local strain can be accommodated. The structural symmetry of S-K QDs is therefore intrinsically linked to the strain symmetry of the crystal substrate, in addition to subsequent preferential crystal growth in the [110] direction, and underlying rotational asymmetry of the atomic surface.

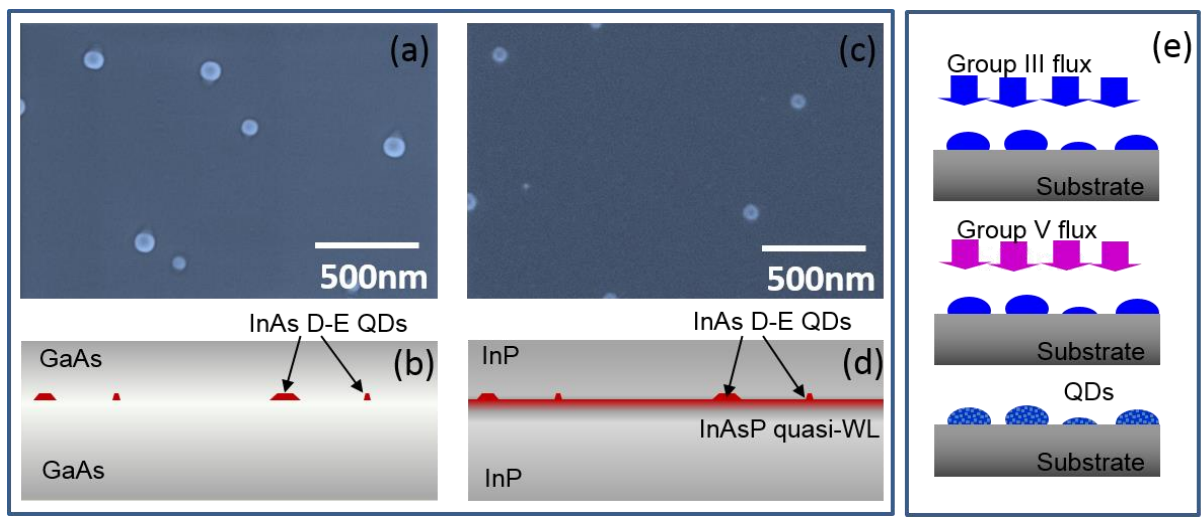

Figure 1: Illustration of droplet formation. Scanning electron micrographs of uncrystallised indium droplets deposited on GaAs (001) (a) and InP (001) surface (c). (b) and (d) are cross section cartoons of the D-E QDs grown in GaAs and InP respectively. The droplet epitaxy growth scheme is illustrated in (e): a substrate is deposited with group III material followed by crystallisation with group V flux leading to formation of quantum dots.

Droplet epitaxy on the other hand does not rely on strain during QD formation. Indium and arsenic fluxes are supplied separately at lower temperature compared to S-K QDs (see Figure 1(e) for illustration). The QDs are formed first through the creation of liquid indium droplets on the substrate, followed by reaction with arsenic to form crystalline structures. Formation of a wetting layer is therefore avoided. Droplet epitaxy has been employed in strain free systems, such as GaAs/AlAs QDs where formation of high quality 2D layers is difficult, especially on (111) crystal surfaces. D-E QDs grown on (001) surfaces have so far been limited, even though it has been suggested that the liquid nature of the droplet produces dislocation free coherent nanocrystals[16].

Examples of indium droplets on GaAs and InP surfaces are shown in Figure 1 (a) and (c). The growth technique is presented in Figure 1(e). Independent of the surface grown on, indium droplets are round and symmetric. The droplets diameter varies from 60 to 100 nm in the case of the GaAs substrate, while the dots on the InP surface tend to be smaller and of diameter less than 60nm. The droplet density is $2\text{-}4\cdot10^8\,\text{cm}^{-2}$ in both cases. As the droplet epitaxy technique is not strain driven, D-E QDs formed on GaAs do not form a wetting layer typical for S-K QDs. However, since droplets are deposited on InP and crystallised with arsenic, a so called "quasi wetting layer" is formed as a result of group V element intermixing. This process cannot be avoided for this material system. This is indicated in Figure 1 (b) and (d).

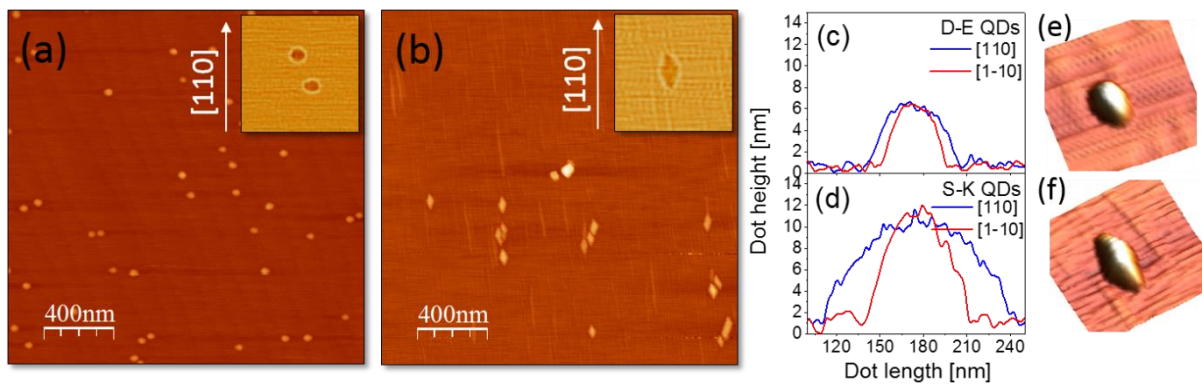

Figure 2: Comparison of uncapped S-K and D-E dots grown on InP. Atomic force microscope (AFM) images of metalorganic vapour phase epitaxy (MOVPE) crystallised D-E QDs on InP (001) surface (a) and S-K QDs on InP (001) surface (b). The scan area in both cases is $2\text{x}2\mu\text{m}^2$. The insets are zoomed in images of individual dots. Line scans over a typical D-E QD and S-K QD on InP are shown on (c) and (d) respectively. (e) and (f) are 3D representations of AFM scans of D-E QD and S-K QD on InP respectively.

A comparison of uncapped S-K and D-E quantum dots grown on InP (001) was characterised by AFM analysis (Figure 2). A number of dot AFM images were fitted with ellipses for statistical analysis of the dot aspect ratio ($R_x/R_y$) and angle of elongation.

The crystallisation process of Indium droplets into InAs QDs does not influence the circular symmetry of the dots significantly as illustrated by the MOVPE grown D-E QDs crystallized on the surface of InP shown in Figure 2(a). The D-E QDs remain symmetric with a mean in-plane aspect ratio of 0.91. An atomic force microscope (AFM) image of similar S-K dots grown by MOVPE on InP is shown in Figure 2(b). Here, the dots have rhombic base, elongated towards [110]. The mean value of their aspect ratio is 0.53. Moreover, pronounced dashes can also be observed. Line scans over a typical D-E and S-K QDs are presented in Figure 2(c) and (d) showing the difference in the dot profile along orthogonal crystallographic directions, which is pronounced for S-K QDs. This is also observed in corresponding AFM images of individual D-E and S-K QDs shown in Figure 2(e) and (f).

One of the chief advantages of the (001) surface is the well-established growth of high quality optical microcavities. Such optical structures are used to greatly improve the collection efficiency of light emitted by QDs, and are widely utilised in single QD optical experiments.

Short wavelength D-E QDs were grown by MBE on GaAs (001) substrates. 1.4ML of In was deposited at 390 °C on Ga terminated GaAs buffer. The crystallization process involved supplying $As_4$ to the growth chamber and ramping the temperature to 500 °C followed by capping the dots with GaAs. Figure 1a shows In droplets deposited at 390 °C. The substrate was immediately cooled down for surface characterisation.

MOVPE grown D-E QDs on InP were formed by deposition of 2ML of In droplets on InP surface via pyrolysis of trimethylindium at 400 °C while withholding the supply of arsine to the growth chamber. Dot crystallization process under arsine overpressure started at 400 and carried on until the substrate reached 500 °C. Next the dots were capped with 30nm of InP followed by more InP at 640 °C. For surface dot characterisation the growth process was stopped after In deposition (Figure 1b) or after droplet crystallization (Figure 2b).

S-K growth by MOVPE on InP was performed at a temperature of 500 °C. A low growth rate of 0.05 nm/s was employed while restricting the nominal thickness of InAs to 2 ML promoted a low density QD growth. InP overgrowth was performed in two stages: first 0.05 nm/s growth rate at 500 °C followed by 0.43 nm/s growth rate at 640 °C.

In all three cases of dot formation described above dot density less than $10^9 cm^{-2}$ was achieved.

The telecom device comprised of 20 pair of DBR mirror, each consisting of 112nm of $(Al_{.15}Ga_{.75})_{.48}In_{.52}As$ and 123nm of InP followed by 5/4 lambda InP and dots capped with 6/4 lambda InP. The short wavelength device consisted of 12 pair bottom DBR mirror, each consisting of 65nm of GaAs and 77nm of AlAs followed by 2 lambda GaAs cavity comprising dots in the middle. The cavity was finished off with two repeats of top mirror.

Low temperature micro-photoluminescence (µPL) measurements were performed with a 785 nm CW excitation laser and the samples mounted in a liquid helium-cooled cryostat. Emission was coupled into single mode fibre to isolate signal from individual quantum dots. Sending the fibre-coupled light to a spectrometer equipped with a silicon CCD camera or InGaAs photodiode array, we were able to perform spectrally resolved measurements.

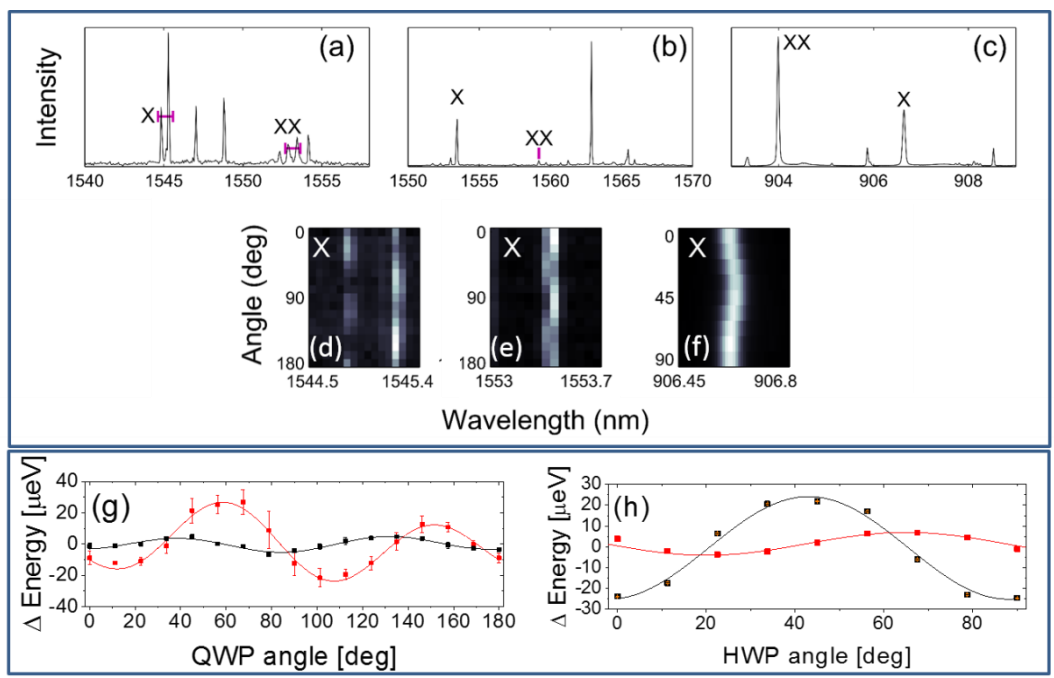

Figure 3: (a-c) Spectra of individual InAs quantum dots for (a) S-K QDs on InP emitting at ~1550 nm (b) D-E QDs on InP emitting at ~1550 nm, and (c) S-K QDs on GaAs emitting at ~900 nm. X and XX transitions are identified. (d-f) Corresponding polarised X emission spectra as a function of the angle of a quarter-wave plate (d-e) or half-wave plate (f). (g-h) Examples FSS measurements of (g) D-E QDs on InP emitting at ~1550 nm, and (h) D-E QDs on GaAs emitting at ~900 nm. Each panel comprises two example measurements for X lines with small and large FSS. For details see text.

FSS measurements were performed by sending the photoluminescence (PL) light from a single dot through a rotatable wave plate and fixed linear polariser, and measuring the resulting shift in energy at the spectrometer. We employ two variations of this technique. The first uses a half-wave plate (HWP), which is the established technique to measure FSS of quantum dots[5,7]. The second method uses a quarter-wave plate (QWP), which we show can extract information on the FSS, the orientation of the quantum dot's dipole moment with respect to the lab frame, and the birefringence accumulated in optical fibre (Appendix). This differs from the traditional half-wave plate technique, which provides just the FSS and dipole orientation, limiting it to free space applications.

In our experiments, we employ the QWP method for InAs/InP D-E QDs, on light emerging from single mode fibre, and the HWP method for InAs/GaAs D-E QDs, in free space before coupling into fibre. Where the FSS was below the resolution of the spectrometer, we could observe shifts by performing fits to the line shape, providing an accuracy of ~2 μeV.

Figure 3 shows the spectra corresponding to a small number of QDs emitting in the cavity mode for (a) S-K QDs on InP emitting around 1550 nm, (b) D-E QDs on InP emitting around 1550 nm, and (c) D-E QDs on GaAs emitting around 900nm. . The excitonic transitions were identified by power dependence and FSS measurements. The plots (d-f) below each panel (a-c) indicate the spectral shifts on the exciton (X) line in each spectrum when filtering emission for different polarisations, as a function of the waveplate angle.  The FSS value for the X line in Figure 3(a) is 235±3 μeV, which is typical for S-K dots at ~1550 nm wavelengths. FSS measured on D-E QDs is significantly smaller: 29±1 μeV around 1550 nm in Figure 3(b) and 31±1 μeV for D-E QDs at around 900 nm in Figure 3(c).

Example FSS measurements are shown in Figure 3(g) and (h) for DE-QDs on InP and D-E QDs on GaAs, respectively. Energy shifts as a function of waveplate angle are well resolved, and fit well with expected behaviour (solid lines). More complex dependence is observed for the QWP method, due to presence of richer information compared to the HWP method. For the examples shown here, we measure 17 μeV, and 79 μeV for the FSS of D-E QDs on InP, and 10 μeV and 49 μeV for the FSS of D-E QDs on GaAs.

Figure 4 summarises the statistics of FSS taken on both D-E and S-K QDs grown on InP and GaAs. The distribution of FSS for S-K QDs on InP emitting around 1550 nm is shown in Fig. 4(a). The FSS is found to be very large, with measured mean value of 176±9 μeV. This is attributed to highly elongated quantum dots[17], confirmed by the mean aspect ratio of 0.53 measured by AFM for uncapped S-K QDs on InP. For such large FSS, significant reduction through annealing[18] or use of external magnetic[19], electric[20,21], or strain fields[22,23] is impractical.

The difference between S-K dots and D-E QDs grown on InP is substantial. The FSS measured for D-E QDs is more than 4x smaller and has a mean value of 42±2 µeV (see Figure 4 (b)), attributed to the more symmetric base aspect ratio of 0.91. Also the distribution of FSS is significantly smaller for droplets compared to S-K dots, with a standard deviation of 17.7 µeV compared to 58.8 µeV for S-K dots.

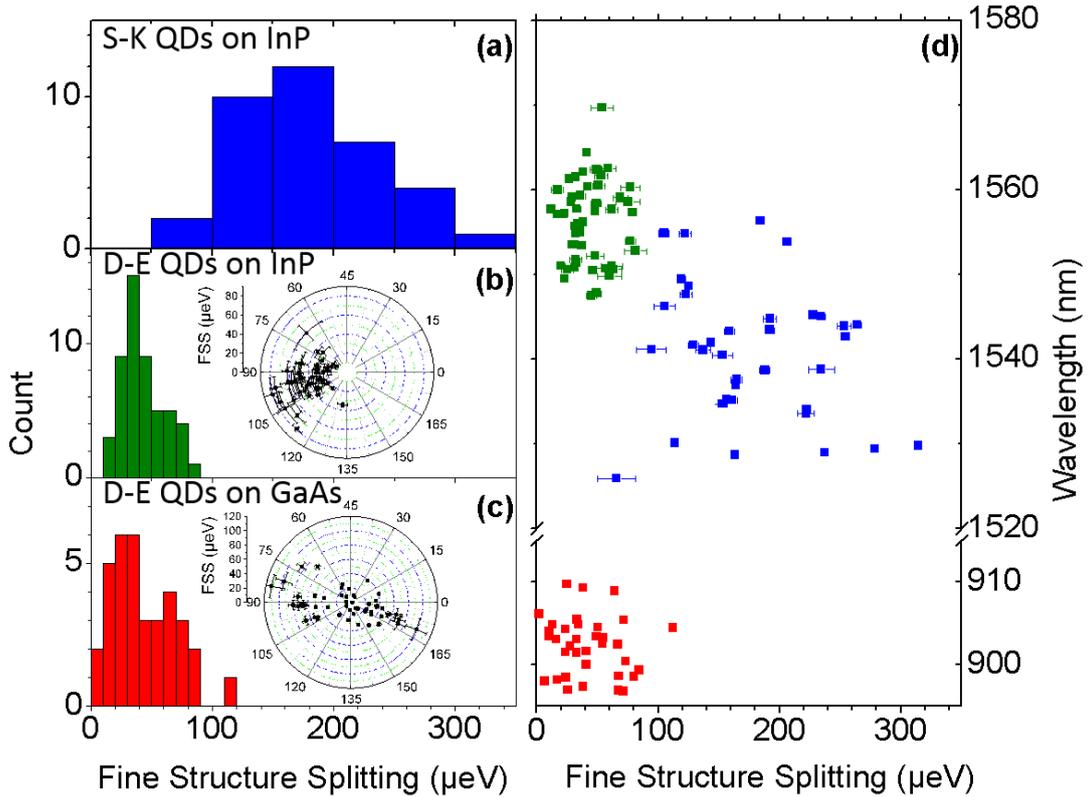

Figure 4: Exciton fine structure splitting of quantum dots emitting at around 900 nm and 1550 nm (telecom C-band). (a) FSS histogram for S-K QDs grown on InP emitting at 1550nm (b) FSS histogram for D-E QDs grown on InP emitting around 1550nm. (c) FSS histogram for D-E QDs grown on GaAs emitting around 900 nm. (d) Corresponding FSS dependence on the emission wavelength measured for D-E (green) and S-K (blue) QDs grown on InP, and D-E QDs grown on GaAs (red). The insets in (b) and (c) show polar plots for FSS values measured for D-E QDs grown on InP and GaAs respectively.

Similar statistical analysis was carried out for D-E QDs dots grown on GaAs emitting around 900 nm (Figure 4(c)). The FSS distribution is remarkably similar to that of Figure 4(b) for D-E QDs grown on InP. The FSS distribution has a mean 42±4 µeV, with a standard deviation 30.5 µeV.

The polar plots in Figure 4 (b) and (c) show the QD dipole orientation measured for D-E QDs grown on InP and GaAs respectively. D-E QDs on InP tend to have dipoles aligned along one privileged direction. D-E QDs on GaAs tend to have a more random dipole distribution, particularly for FSS less than 40 μeV. This may be associated with the lack of a wetting layer and with the fact that larger dots have smaller aspect ratios.

Previous reports have shown correlation between the emission wavelength and the FSS of the quantum dot for both S-K dots[7] and D-E dots[10]. However, for the D-E QDs studied here, we observe no obvious correlation, as shown in Figure 4(d). This has a particular consequence for D-E QDs emitting around 900nm, as it lifts the restriction to operate at the wavelength corresponding to minimum FSS, typically 885 nm. We also observe no dependence on the FSS for S-K QDs with emission wavelength at around 1550 nm, which suggests that the QD shape can vary independently from the size for both growth techniques.

The realisation of quantum dot based entangled light sources for the conventional telecom band requires quantum dots emitting around 1550 nm, with small FSS, and embedded in an optical microcavity. Our results show such quantum dots can be produced using droplet epitaxy of InAs QDs on (001) InP substrates. The reduction in FSS compared to an S-K QD sample grown at similar wavelength is a factor of 4 smaller, and critically the fraction of QDs with less than 40μeV splitting (the mean FSS value for D-E QDs) is increased from 0% to 53%. Furthermore, growth of InAs droplets on GaAs provides a universal growth scheme, allowing QDs to be realised with a similar FSS distribution but emitting at ~600 nm shorter wavelength around 900 nm, within different semiconductor materials. Statistics show that the new scheme improves the symmetry of the dots by 72%. Our study opens up the route to produce low FSS QDs at any wavelengths between 900 nm and 1550 nm. This is of a high importance not only for quantum communications and quantum cryptography but for any quantum dot related research.

**Acknowledgements**

The authors would like to acknowledge partial financial support of UK EPSRC and InnovateUK (project 102245). We also would like to thank Guillaume Bourmaud for writing the AFM analysis software code.

**Additional information**

Correspondence should be addressed to JSS, or AJS.

**Appendix: Measurement of fine-structure-splitting.**

Suppose our quantum dots emit light in the state $\rho_1$, where the H and V polarizations are eigenstates of the Hamiltonian $\hat{H}$,

$$\rho_1 = \left(\frac{1+p}{2}\right)|H\rangle\langle H| + \left(\frac{1-p}{2}\right)|V\rangle\langle V| \quad (1)$$

$$\hat{H}\rho_1 = E_H\left(\frac{1+p}{2}\right)|H\rangle\langle H| + E_V\left(\frac{1-p}{2}\right)|V\rangle\langle V| \quad (2)$$

with energies $E_H$ and $E_V$ respectively. In the case that degeneracy of these states is lifted, they will exhibit a fine structure splitting $s = E_H - E_V$, and the mean energy is given by $\varepsilon = (E_H + E_V)/2$. The state may also have some polarization $p$.

After travelling through a series of optics and fibres, typically birefringent media, the state will have undergone a rotation $\theta$ and phase shift $\varphi$ in its polarization.

$$|H\rangle \rightarrow |B_1\rangle = \cos\frac{\theta}{2}|H\rangle + \sin\frac{\theta}{2}e^{i\phi}|V\rangle \quad (3)$$

$$|V\rangle \rightarrow |B_2\rangle = \sin\frac{\theta}{2}|H\rangle - \cos\frac{\theta}{2}e^{i\phi}|V\rangle \quad (4)$$

$$\rho_1 \rightarrow \rho_2 = \left(\frac{1+p}{2}\right)|B_1\rangle\langle B_1| + \left(\frac{1-p}{2}\right)|B_2\rangle\langle B_2| \quad (5)$$

Immediately before entering our spectrometer, the light is passed through a quarter-wave plate (QWP) at angle $\chi$ to the transmission polarization, that we call H, of a subsequent linear polarizer. (In the case that our measurement H polarization does not precisely correspond to one of the eigenbases from the first equation, we will see some constant offset in θ.) This is equivalent to measuring our state against the measurement basis $|M\rangle$, given by

$$|M(\chi)\rangle = QWP(\chi)|H\rangle = \frac{1}{\sqrt{2}}(i + \cos 2\chi)|H\rangle + \frac{1}{\sqrt{2}}\sin 2\chi |V\rangle \quad (6)$$

The energy we observe at the spectrometer is

$$E(\chi) = \frac{\langle M|\hat{H}\rho_2|M\rangle}{\langle M|\rho_2|M\rangle}$$
$$= \varepsilon + \frac{s}{2}\left(\frac{(\alpha_1-\alpha_2)+p}{1+p(\alpha_1-\alpha_2)}\right) \quad (7)$$

where $\alpha_j = |\langle M|B_j\rangle|^2$. We thereby get an expression for the deviation $\Delta E$ from the mean energy $\varepsilon$ as a function of $\chi$

$$\alpha_1 - \alpha_2 = \frac{1}{2}(\cos\theta\,(1+\cos 4\chi) + \sin\theta\sin 4\chi\cos\phi - 2\sin\theta\sin 2\chi\sin\phi) \quad (8)$$

$$\Delta E(\chi) = E(\chi) - \varepsilon$$
$$= \frac{s}{2}\left(\frac{2p+\cos\theta\,(1+\cos 4\chi)+\sin\theta\sin 4\chi\cos\phi-2\sin\theta\sin 2\chi\sin\phi}{2+p\cos\theta\,(1+\cos 4\chi)+p\sin\theta\sin 4\chi\cos\phi-2p\sin\theta\sin 2\chi\sin\phi}\right) \quad (9)$$

## References

1. O. Benson, C. Santori, M. Pelton, and Y. Yamamoto, "Regulated and Entangled Photons from a Single Quantum Dot," Phys. Rev. Lett. **84**, 2513–2516 (2000).
2. R. M. Stevenson, R. J. Young, P. Atkinson, K. Cooper, D. A. Ritchie, and A. J. Shields, "A semiconductor source of triggered entangled photon pairs," Nature **439**, 179–182 (2006).
3. C. L. Salter, R. M. Stevenson, I. Farrer, C. A. Nicoll, D. A. Ritchie, and A. J. Shields, "An entangled-light-emitting diode," Nature **465**, 594–597 (2010).
4. M. B. Ward, M. C. Dean, R. M. Stevenson, A. J. Bennett, D. Ellis, K. Cooper, I. Farrer, C. A. Nicoll, D. A. Ritchie, and A. J. Shields, "Coherent dynamics of a telecom-wavelength entangled photon source," Nat. Commun. **5, 3315** (2014).
5. D. Gammon, E. S. Snow, B. V. Shanabrook, D. S. Katzer, and D. Park, "Fine structure splitting in the optical spectra of single GaAs quantum dots," Phys. Rev. Lett. **76**, 3005–3008 (1996).
6. V. D. Kulakovskii, G. Bacher, R. Weigand, T. Kümmell, A. Forchel, E. Borovitskaya, K. Leonardi, and D. Hommel, "Fine Structure of Biexciton Emission in Symmetric and Asymmetric CdSe/ZnSe Single Quantum Dots," Phys. Rev. Lett. **82**, 1780–1783 (1999).
7. R. J. Young, R. M. Stevenson, A. J. Shields, P. Atkinson, K. Cooper, D. A. Ritchie, K. M. Groom, A. I. Tartakovskii, and M. S. Skolnick, "Inversion of exciton level splitting in quantum dots," Phys. Rev. B **72**, 113305 (2005).
8. G. Juska, V. Dimastrodonato, L. O. Mereni, A. Gocalinska, and E. Pelucchi, "Towards quantum-dot arrays of entangled photon emitters," Nat. Photon. **7**, 527–531 (2013).
9. T. Kuroda, T. Mano, N. Ha, H. Nakajima, H. Kumano, B. Urbaszek, M. Jo, M. Abbarchi, Y. Sakuma, K. Sakoda, I. Suemune, X. Marie, and T. Amand, "Symmetric quantum dots as efficient sources of highly entangled photons. Violation of Bell's inequality without spectral and temporal filtering," Phys. Rev. B **88**, 041306 (2013).
10. X. Liu, N. Ha, H. Nakajima, T. Mano, T. Kuroda, B. Urbaszek, H. Kumano, I. Suemune, Y. Sakuma, and K. Sakoda, "Vanishing fine-structure splittings in telecommunication-wavelength quantum dots grown on (111)A surfaces by droplet epitaxy," Phys. Rev. B **90**, 081301 (2014).
11. T. Miyazawa, K. Takemoto, Y. Sakuma, S. Hirose, T. Usuki, N. Yokoyama, M. Takatsu, and Y. Arakawa, "Single-Photon Generation in the 1.55-μm Optical-Fiber Band from an InAs/InP Quantum Dot," Japanese Journal of Applied Physics **44**, L620 (2005).
12. Ł. Dusanowski, M. Syperek, P. Mrowiński, W. Rudno-Rudziński, J. Misiewicz, A. Somers, S. Höfling, M. Kamp, J. P. Reithmaier, and G. Sęk, "Single photon emission at 1.55 μm from charged and neutral exciton confined in a single quantum dash," Appl. Phys. Lett. **105**, 21909 (2014).
13. M. D. Birowosuto, H. Sumikura, S. Matsuo, H. Taniyama, P. J. van Veldhoven, R. Nötzel, and M. Notomi, "Fast Purcell-enhanced single photon source in 1,550-nm telecom band from a resonant quantum dot-cavity coupling," Sci. Rep. **2**, 321 (2012).
14. K. Takemoto, Y. Sakuma, S. Hirose, T. Usuki, and N. Yokoyama, "Observation of Exciton Transition in 1.3–1.55 μm Band from Single InAs/InP Quantum Dots in Mesa Structure," Japanese Journal of Applied Physics **43**, L349 (2004).
15. S. Anantathanasarn, R. Nötzel, P. J. van Veldhoven, T. J. Eijkemans, and J. H. Wolter, "Wavelength-tunable (1.55-μm region) InAs quantum dots in InGaAsP/InP (100) grown by metal-organic vapor-phase epitaxy," Journal of Applied Physics **98**, 013503 (2005).
16. T. Mano, K. Watanabe, S. Tsukamoto, H. Fujioka, M. Oshima, and N. Koguchi, "New Self-Organized Growth Method for InGaAs Quantum Dots on GaAs(001) Using Droplet Epitaxy," Japanese Journal of Applied Physics **38**, L1009 (1999).
17. L. He, M. Gong, C.-F. Li, G.-C. Guo, and A. Zunger, "Highly Reduced Fine-Structure Splitting in InAs/InP Quantum Dots Offering an Efficient On-Demand Entangled 1.55-μm Photon Emitter," Phys. Rev. Lett. **101**, 157405 (2008).
18. D. J. P. Ellis, R. M. Stevenson, R. J. Young, A. J. Shields, P. Atkinson, and D. A. Ritchie, "Control of fine-structure splitting of individual InAs quantum dots by rapid thermal annealing," Appl. Phys. Lett. **90**, 11907 (2007).
19. R. M. Stevenson, R. J. Young, P. See, D. G. Gevaux, K. Cooper, P. Atkinson, I. Farrer, D. A. Ritchie, and A. J. Shields, "Magnetic-field-induced reduction of the exciton polarization splitting in InAs quantum dots," Phys. Rev. B **73**, 33306 (2006).
20. K. Kowalik, O. Krebs, A. Lemaître, B. Eble, A. Kudelski, P. Voisin, S. Seidl, and J. A. Gaj, "Monitoring electrically driven cancellation of exciton fine structure in a semiconductor quantum dot by optical orientation," Appl. Phys. Lett. **91**, 183104 (2007).
21. A. J. Bennett, M. A. Pooley, R. M. Stevenson, M. B. Ward, R. B. Patel, de la Giroday, A. Boyer, N. Sköld, I. Farrer, C. A. Nicoll, D. A. Ritchie, and A. J. Shields, "Electric-field-induced coherent coupling of the exciton states in a single quantum dot," Nat. Phys. **6**, 947–950 (2010).
22. S. Seidl, M. Kroner, A. Högele, K. Karrai, R. J. Warburton, A. Badolato, and P. M. Petroff, "Effect of uniaxial stress on excitons in a self-assembled quantum dot," Appl. Phys. Lett. **88**, 203113 (2006).
23. J. Zhang, J. S. Wildmann, F. Ding, R. Trotta, Y. Huo, E. Zallo, D. Huber, A. Rastelli, and O. G. Schmidt, "High yield and ultrafast sources of electrically triggered entangled-photon pairs based on strain-tunable quantum dots," Nat. Commun. **6**, 10067 (2015).